\documentclass[prd, aps, superscriptaddress, preprintnumbers, twocolumn, floatfix, nofootinbib]{revtex4}
\pdfoutput=1

\usepackage{amsfonts}
\usepackage{amsmath}
\usepackage{amssymb}
\usepackage{bm}
\usepackage{dcolumn}
\usepackage{graphicx}   
\usepackage[latin1]{inputenc}
\usepackage{latexsym}
\usepackage{rotating}
\usepackage{hyperref}
\usepackage{graphicx}
\usepackage{color}

\newcommand\be{\begin{equation}}
\newcommand\ba{\begin{eqnarray}}
\newcommand\ee{\end{equation}}
\newcommand\ea{\end{eqnarray}}

\begin{document}

\title{Relaxing the TCC Bound on Inflationary Cosmology?}

\author{Vahid Kamali}
\email{vkamali@ipm.ir}
\affiliation{Department of Physics, McGill University, Montr\'{e}al, QC, H3A 2T8, Canada\\
Department of Physics, Bu-Ali Sina (Avicenna) University, Hamedan 65178, 016016, Iran\\
School of Physics, Insitute for Research in Fundamental Sciences (IPM), 19538-33511, Tahran, Iran}

\author{Robert Brandenberger}
\email{rhb@physics.mcgill.ca}
\affiliation{Department of Physics, McGill University, Montr\'{e}al, QC, H3A 2T8, Canada}

\date{\today}

\begin{abstract}

We demonstrate that the strict upper bounds on the energy scale of inflation and on the tensor-to-scalar ratio can be somewhat relaxed if we assume that - after an initial period of slow rolling when scales probed today in CMB experiments exit the Hubble radius - the equation of state of the background changes to correspond to an almost marginally accelerating universe. Constructing an actual model in which this happens appears, however, to be unnatural.

\end{abstract}

\pacs{98.80.Cq}
\maketitle

\section{Introduction} 
\label{sec:intro}

The inflationary scenario, the current paradigm of early universe cosmology, is currently facing serious challenges coming from fundamental physics. On one hand, there are {\it swampland} criteria (see e.g. \cite{swamp} for original works and \cite{swamprev} for reviews) which imply that the potential energy functions required to obtain inflation in an effective field theory of simple \footnote{Note that these constraints are alleviated \cite{warm2} in models of {\it warm inflation} \cite{warm}.} single field inflation is inconsistent with constraints coming from superstring theory (see e.g. \cite{noinflation}). On the other hand, a generalization of Penrose's cosmic censorship \cite{Penrose} has been proposed \cite{BV} which states that scales which were trans-Planckian in the early universe must remain hidden by a Hubble horizon \footnote{This also shields (at least at the level of linear fluctuations) super-Hubble scales from the non-unitarity of an effective field theory analysis \cite{Weiss}.}. This is called the {\it Trans-Planckian Censorship Conjecture} (TCC). Assuming standard vacuum initial conditions for the fluctuations, and standard Big Bang cosmology after the end of inflation, the TCC leads to severe constraints on any inflationary model \cite{BBLV}. Demanding that inflation be consistent with the TCC and allow for a causal structure formation scenario leads to an upper bound
\be \label{Vbound}
V_{e}^{1/4} \, < \, 10^{10} {\rm GeV} \, ,
\ee
where $V_{e}$ is the potential energy density at the end $t = t_e$ of the period of inflation. Assuming that the potential energy does not change by a large factor between the time $t_*$ when scales which are currently observated exit the Hubble radius and the end of inflation \footnote{This is a well justified approximation in simple models of inflation.} this leads to an upper bound on the tensor to (observed) scalar ratio $r$ of \footnote{Note that these bounds can in fact be strengthened in the case that the inflationary epoch is preceded by a radiation phase with begins at the Planck scale \cite{Edward}.}
\be \label{Rbound}
r \, < \, 10^{-30} \, .
\ee

Some ways to mitigate these constraints have been suggested, e.g. by assuming a non-standard cosmological history after the end of inflation \cite{mod1}, non-standard initial conditions for the fluctuations \cite{mod2}, and others \cite{mod3} (see also \cite{mod4}) \footnote{On the other hand, if we add the assumption that inflation is generated by a potential of reasonable shape which yields a spectrum of cosmological perturbations in agreement with the CMB data, the constraints on the scale of inflation are even more stringent \cite{Kadota}.}. Here, we explore by how much the above constraints can be relaxed by allowing for the largest possible running of the potential energy between $t_*$ and $t_e$. Note that given the observational constraint on the slope and running of the spectrum of cosmological perturbations on scales which are currently being probed, it will require severe tunings of the potential in order to be able to obtain the maximal change in the potential energy which we here discuss. A similar study to the one we report on was done in \cite{Mizuno}, who also asked how much the constraints can be relaxed by letting the equation of state during inflation approach the limiting value which one requires for accelerated expansion. Whereas our analysis makes use of a specific model for inflation, the one by \cite{Mizuno} is more general. The bounds we obtain on the tensor to scalar ratio are similar, though in \cite{Mizuno} they depend on the reheating scale.

A few words on notation: We work with the usual Friedmann-Robertson-Walker metric given by the line element
\be
ds^2 \, = \, dt^2 - a(t)^2 d{\bf x}^2 \, ,
\ee
where $t$ is physical time, ${\bf x}$ are the comoving spatial coordinates and $a(t)$ is the cosmological scale factor. The Hubble expansion rate is $H(t)$ whose inverse is the Hubble horizon, and the Planck mass (whose inverse is the Planck length $l_{pl}$) is denoted by $m_{pl}$. We work in natural units in which the speed of light and Planck's constant are set to $1$.

\section{The TCC Bound for Inflationary Cosmology}

The mathematical form of the TCC \cite{BV} applied to inflationary cosmology is \cite{BBLV}
\be \label{TCC}
\frac{a(t_e)}{a(t_i)} l_{pl} \, \leq \, H^{-1}(t_e) \, ,
\ee
where $t_i$ is the beginning of inflation, and $t_e$ is the end of the inflationary phase. This expresses the condition that no length which were sub-Planckian at the beginning of inflation crossed the Hubble horizon by the end of inflation. After inflation ends and the universe no longer undergoes accelerated expansion, no wavelengths ever exit the Hubble radius, and hence the condition (\ref{TCC}) is sufficient for all times.

The reason why the Hubble radius is the relevant scale is the following: in terms of the canonical fluctuation variables (those whose action has a canonical kinetic term), the fluctuations oscillate on sub-Hubble scales. It is only on super-Hubble scales that the fluctuations become squeezed (see \cite{MFB} for a comprehensive review of the theory of cosmological perturbations and \cite{RHBfluctsrev} for an overview) and classicalize \cite{Kiefer}. Hence, if we do not want the classical fluctuations to depend on trans-Planckian physics, we do not want any trans-Planckian modes to cross the Hubble radius.  

Note that (\ref{TCC}) is a necessary but not necessarily a sufficient condition for predictions in cosmology to be shielded from trans-Planckian effects \footnote{Note that the sensitivity of the predictions of standard inflationary models to trans-Planckian physics was suggested a long time ago in \cite{Jerome} (see e.g. \cite{JeromeRev} for a review and references to the large volume of followup works). However, if one assumes that local Lorentz invariance holds down to the smallest scales, then it can be shown that there is no such problem \cite{ST}.}.

The weakest possible form of the TCC (still assuming constant Hubble expansion rate during inflation, standard cosmology after the end of inflation, and the usual initial conditions for fluctuations) results from taking the initial time of inflation $t_i$ to be the time $t_*$ when the current comoving Hubble scale exits the Hubble radius, i.e. by having inflation last the least number of e-foldings consistent with inflation providing a causal mechanism of structure formation. Thus, (\ref{TCC}) becomes
\be \label{TCC2}
\frac{a(t_e)}{a(t_*)} l_{pl} \, \leq \, H^{-1}(t_e) \, .
\ee
In the following we will use the abbreviated notation $a_e \equiv a(t_e)$ and $a_* \equiv a(t_*)$.

The condition (\ref{TCC2}) yield the upper bound (\ref{Rbound}) on the energy scale of inflation. This is easy to understand: the TCC leads to an upper bound on the duration of inflation, but demanding that the current Hubble scale was smaller than the Hubble radius at the beginning of inflation yields a lower bound. The smaller the energy scale of inflation, the more e-foldings of exponential expansion it takes for the Planck length to expand to the Hubble length, and in this way the TCC can be satisfied for low-scale inflation. Given the upper bound on the energy scale of inflation, an upper bound on the amplitude of the spectrum of gravitational waves results, given by (\ref{Rbound}).

Note that a specific inflationary models which are safe from the TCC constraint have been proposed in \cite{safe}. Applications of the TCC to other cosmological issues have been considered in \cite{other}. The implications of the TCC for holographic cosmology have been studied in \cite{Bernardo}, and connections between the TCC and the swampland criteria have been explored in \cite{connection}.

\section{Relaxing the TCC Bound}

In this Letter we investigate to what extent the above bounds can be relaxed by abandoning the assumption that the Hubble expansion rate is fixed during inflation. We will be working in the context of inflation mediated by a canonically normalized scalar field $\varphi$ slowly rolling in a potential $V(\varphi)$. In order to obtain a small scalar spectrum tilt on observable scales, the potential has to change very little during the couple of e-foldings of inflation when the range of scales which are currently probed by CMB and large-scale structure observations exit the Hubble radius. But it is possible to imagine that after these scales have exited the Hubble horizon, the field starts to roll faster, being only constrained by the requirement that accelerated expansion persists, i.e. that
\be
{\dot{\varphi}}^2 \, < \, V(\varphi) \, .
\ee
We will in the following assume that the ratio of kinetic and potential energy is independent of time (after the initial period when scales which are observed today exit the Hubble radius), i.e. we take
\be \label{betadef}
{\dot{\varphi}}^2 \, = \, \beta V(\varphi) \, ,
\ee
where $\beta$ is a constant which must satisfy
\be
\beta \, < \, 1 
\ee
in order to have accelerated expansion. This model is called {\it constant roll inflation} \cite{roll1} and its predictions have been compared to observations in \cite{roll2}.

In order to be consistent with the TCC, there is an upper bound on the duration of inflation and hence an upper bound on the change in the potential energy between the time $t_*$ when the current comoving Hubble scale exits the Hubble radius (we will denote with $V_*$ the value of the potential energy density at this time) and the end of inflation when the potential energy is $V_e$. Assuming that the inequality (\ref{TCC2}) is saturated, we can derive (see Appendix) the following relation between $V_*$ and $V_e$:
\be \label{Vratio}
V_e \, = \, \bigl( \frac{1}{3} (1 + \frac{\beta}{2}) \frac{V_e}{m_{pl}^4} \bigr)^{\tilde{\beta}} V_* \, ,
\ee
with
\be
{\tilde{\beta}} \, \equiv \, \frac{3\beta}{2(1 + \beta/2)} \, .
\ee
Taking all the dependence on $V_e$ to the left hand side of the equation we get 
\be 
V_e^{1 - \tilde{\beta}} \, = \, \kappa m_{pl}^{- 4 \tilde{\beta}} V_* \, ,
\ee
with
\be
\kappa \, \equiv \, \bigl( \frac{1}{3} (1 + \frac{\beta}{2}) \bigr)^{\tilde{\beta}} \, .
\ee

The condition that the comoving length corresponding to the current Hubble radius $H_0^{-1}$ exits the Hubble radius after time $t_*$ reads
\be \label{SF}
H_0^{-1} \frac{a_*}{a_e} \frac{a_e}{a_0} \, < \, H_*^{-1} \, ,
\ee
where the subscripts $0$ stands for the present time.

We will assume that the Standard Big Bang phase of radiation domination begins after the end of inflation. In this case the ratio of the scale factors $a_e$ and $a_0$ is the inverse of the corresponding radiation temperatures (if we neglect entropy production between $t_e$ and $t_0$ (taking entropy production into account would only change our result by a ${\cal{O}}(1)$ factor). Thus
\be \label{Aratio}
\frac{a_e}{a_0} \, = \, \frac{T_0}{T_e} \, = \, 3^{-1/4} g^*(T_e)^{1/4} \frac{T_0}{(m_{pl} H_e)^{1/2}} \, ,
\ee
where we have used
\be
g^*(T_e) T_e^4 \, = \, V_e \, 
\ee
and the Friedmann equation which relates $V_e$ and $H_e$. Note also that $T_0$ and $H_0$ are related via
\be
g^*(T_0) T_0^4 \, = \, \frac{T_0}{T_{eq}} \rho_0 \, = \, 3 m_{pl}^2 H_0^2 \frac{T_0}{T_{eq}} \, ,
\ee
where $T_{eq}$ is the temperature at the time of equal matter and radiation. In the above, $g^*$ is the number of effective spin degrees of freedom in the radiation bath.
  
Let us now return to the condition (\ref{SF}) for inflation to last sufficiently long to be able to provide a causal mechanism for the generation of structure. Assuming that the TCC is saturated yields
\be \label{Bratio}
 \frac{a_e}{a_*} \, = \, \frac{m_{pl}}{H_e}  \, .
\ee
Inserting (\ref{Aratio}) into (\ref{Bratio}) into (\ref{SF}) and making use of (\ref{Vratio}) yields 
\ba
H_* \, & < & \, H_0 \frac{a_0}{a_e} \frac{a_e}{a_*} \\
&=& \, g^*(T_0)^{1/2} g^*(T_e)^{-1/4} \frac{T_0 m_{pl}}{V_e^{1/4}} \bigl( \frac{T_{eq}}{T_0} \bigr)^{1/2}
(1 + \beta/2)^{-1/4} \, .
\nonumber
\ea
We now use the Friedmann equation to relate $H_*$ and $V_*$
\be
H_*^2 m_{pl}^2 \, = \, \frac{1}{3} V_* (1 + \beta/2) \, 
\ee
and insert the relation (\ref{Vratio}) between $V_e$ and $V_*$ to turn the above inequality into an upper bound on the energy scale $V_*$ of inflation
\be \label{Vbound}
V_*^{1/4} \, < \, {\tilde{\kappa}} m_{pl} \bigl( \frac{T_0}{m_{pl}} \bigr)^{\gamma} \, .
\ee
with 
\be
{\tilde{\kappa}} \, \equiv \, \bigl( \frac{1}{1 + \beta/2} \bigr)^{\gamma}
3^{\gamma/2} (g^*(T_0))^{\gamma/2} (g^*(T_e))^{- \gamma/4}
\bigl( \frac{T_{eq}}{T_0} \bigr)^{\gamma/2} \kappa^{- \gamma / (4(1 - \tilde{\beta}))}
\ee
where 
\be
\gamma \, = \frac{1}{3 + {\tilde{\beta}}/(1 - {\tilde{\beta}})} \, .
\ee
Note that ${\tilde{\kappa}}$ is a factor of order unity.

To evaluate the above bound, we will consider the case
\be
\tilde{\beta} = \, \frac{3}{4} \, .
\ee
This corresponds to a value of $\beta$ close to its upper bound.
In this case, neglecting factors of order unity, the bound (\ref{Vbound}) becomes
\be \label{scale}
V_{*}^{1/4} \, < \, m_{pl} \bigl( \frac{T_0}{m_{pl}} \bigr)^{1/6} \, \sim \, 10^{-5} m_{pl} \, 
\ee
which is lower than the scale of inflation in the simplest slow-roll models, but much higher than the scale determined using an almost constant value of $H$ during inflation.

The energy scale of inflation when scales observed today exit the Hubble radius determines the amplitude of the gravitational wave spectrum on these scales. The power spectrum of gravitational waves is
\be
{\cal{P}}_h \, \sim \, \bigl( \frac{H}{m_{pl}} \bigr)^2 \, \sim \frac{V_*}{m_{pl}^4} \, < \, 10^{-20} \, .
\ee
Assuming that the scalar fluctuations have the amplitude which is observed, the resulting tensor to scalar ratio $r$ obeys the bound
\be \label{rbound}
r \, < \, 10^{-10} \, ,
\ee
which is far lower than the value which will be measurable in upcoming CMB polarization experiments.

\section{Conclusions and Discussion} \label{conclusion}

We have studied to what extent the upper bound on the energy scale of inflation which follows from the TCC can be relaxed by removing the condition that $H$ is approximately constant during the entire period of inflation. We have shown that it is possible to relax the upper bound on the scale of inflation to the value given in (\ref{scale}), which relaxes the upper bound on the tensor to scalar ratio to (\ref{rbound}). This is obtained by assuming that after the couple of e-foldings of slow-roll inflation (which is required in order to obtain the small tilte in the scalar spectrum which is observed on cosmological scales), the equation of state of the field driving inflation changes dramatically from $w \simeq -1$ to $w \sim -1/3$, the maximal value which is consistent with having accelerated expansion of space.

Note that the bounds (\ref{scale}) and (\ref{rbound}) are not strict upper bounds. They are obtained for a value of the equation of state which is close to the upper bound, but which does not quite saturate it. It is important, however, to emphasize that the more the equation of state deviates from $w \simeq -1$, while maintaining slow-roll when scale of cosmological interest exit the Hubble radius, the more baroque the required model building becomes. In fact, a more natural way to raise the upper bound on the tensor to scalar ratio $r$ is by considering warm inflation \cite{warm}. In this context, a Hubble constant which decreases substantially during inflation is naturally obtained, and hence a larger value of $r$ can be obtained for similar reasons to the ones we have discussed here. This was worked out recently in \cite{mod3}.

Our results are in good agreement with thos of \cite{Mizuno}, who also considered the extent to which the constraints on the energy scale at the onset of inflation by assuming an equation of state during the inflationary phase which approaches $w = - 1/3$. Our analysis is in the context of a specific model for inflation, namely {\it constant roll inflation}, whereas the one by \cite{Mizuno} is more general. The bounds we obtain on the tensor to scalar ratio are similar, though in \cite{Mizuno} they depend on the reheating scale.

\section*{Acknowledgement}

\noindent The research at McGill is supported in part by funds from NSERC and from the Canada Research Chair program. The visit of VK to McGill has also been supported in part by the McGill Space Institute.

\section*{Appendix}

In this Appendix we derive the key equation (\ref{Vratio}) which is used in the above analysis. Our starting point is (\ref{betadef})
\be
(\dot{\varphi})^2 \, = \beta V \, ,
\ee
Integrating both sides over time yields
\be \label{phiev}
\int d\varphi \, = \, \sqrt{3\beta} \int V^{1/2} dt \, = \, \frac{1}{{\sqrt{\beta/2 + 1}}} \sqrt{3\beta} m_{pl} \int H dt \, .
\ee
If we integrate over the time interval which saturates the TCC we have
\be
e^{\int H dt} \, = \, \frac{m_{pl}}{H_e} \, .
\ee
Inserting this into (\ref{phiev}) yields
\be \label{range}
\Delta \varphi \, = \, \frac{1}{{\sqrt{\beta/2 + 1}}} \sqrt{3\beta} m_{pl} {\rm{ln}} \bigl( \frac{m_{pl}}{H_e} \bigr) \, .
\ee

Consider now the equation of motion for a homogeneous canonically normalized scalar field $\varphi$ in an expanding background
\be
{\ddot{\varphi}} + 3H{\dot{\varphi}} + V^{\prime} \, = \, 0 \, .
\ee
Taking the time derivative of (\ref{betadef}) yields
\be
{\ddot{\varphi}} \, = \, \frac{\beta}{2} V^{\prime} \, ,
\ee
and, inserting this result back into the equation of motion gives
\be
(\frac{\beta}{2} + 1) \frac{V^{\prime}}{V} + \frac{\sqrt{3\beta} \sqrt{1 + \beta / 2}}{m_{pl}} \, = \, 0 \, ,
\ee
or, equivalently,
\be
\frac{dV}{V} \, = \, - \frac{\sqrt{3\beta}}{\sqrt{1 + \beta/2}} \frac{d\varphi}{m_{pl}} \, .
\ee
After integration this becomes
\be
{\rm{ln}} \frac{V_e}{V_*} \, = \, - \frac{\sqrt{3\beta}}{\sqrt{1 + \beta/2}} \frac{\Delta \varphi}{m_{pl}} \, 
= \, - \frac{3\beta}{1 + \beta/2} {\rm{ln}} \frac{m_{pl}}{H_e} \, ,
\ee
where in the second step we have inserted (\ref{range}). After exponentiation, and using the Friedmann equation to substitute for $H_e$ in terms of $V_e$ we obtain (\ref{Vratio}).

\end{document}